%% file: main.tex
\documentclass[aps,prl,showpacs,showkeys,amsmath,amssymb,
twocolumn,
floatfix,
superscriptaddress
]{revtex4-1}
\usepackage{graphicx}
\usepackage{times}
\usepackage{verbatim}
\usepackage{color}
\usepackage{bm}
\usepackage{natbib}
\usepackage[normalem]{ulem}
\definecolor{gray}{rgb}{0.05,0.05,0.55}
\usepackage[colorlinks,linkcolor=gray,anchorcolor=gray,citecolor=gray, urlcolor=gray,filecolor=gray]{hyperref}
\usepackage{array}

\newcommand{\ep}{E$_{\rm p}$}
\newcommand{\enu}{E$_\nu$}

\graphicspath{{figures/}}
\usepackage{graphicx}

\begin{document}


\title{First measurement of high-energy reactor antineutrinos at Daya Bay}

\input{PhysRev}

\date{\today}

\begin{abstract}
This Letter reports the first measurement of high-energy reactor antineutrinos at Daya Bay, with nearly 9000 inverse beta decay candidates in the prompt energy region of 8-12~MeV observed over 1958 days of data collection.
A multivariate analysis is used to separate 2500 signal events from background statistically.
The hypothesis of no reactor antineutrinos with neutrino energy above 10~MeV is rejected with a significance of 6.2 standard deviations.
A 29\% antineutrino flux deficit in the prompt energy region of 8-11~MeV is observed compared to a recent model prediction.
We provide the unfolded antineutrino spectrum above 7 MeV as a data-based reference for other experiments.
This result provides the first direct observation of the production of antineutrinos from several high-$Q_{\beta}$ isotopes in commercial reactors.

\end{abstract}

\maketitle

Recently, antineutrino measurements at commercial~\cite{Mention:2011rk,Adey:2018qct,An:2015nua,An:2016srz,RENO:2015ksa,Abe:2014bwa} and research~\cite{Ashenfelter:2018jrx,Andriamirado:2020erz,AlmazanMolina:2020spe,AlmazanMolina:2020jlh,Prospect:2021lbs} reactors have challenged both the flux and spectral shape of model predictions using either the conversion method~\cite{Schreckenbach:1985ep,VonFeilitzsch:1982jw,Hahn:1989zr,Mueller:2011nm,Huber:2011wv,Haag:2013raa} or the summation method~\cite{Mueller:2011nm,Fallot:2012jv,Dwyer:2014eka,Estienne:2019ujo}.
These discrepancies have spurred substantial advancement of the theory and reexamination of experimental basis of the predictions~, although these improvements have not substantially altered the underlying data-model disagreements~\cite{Kopeikin:2021ugh, Fallot:2012jv,Hayes:2013wra,Estienne:2019ujo,Berryman:2019hme,Bernstein:2019nqq,Hayen:2019eop,Giunti:2021kab}.

In these experiments, antineutrinos are detected via the inverse beta decay (IBD) reaction ($\bar{\nu}_e + p \rightarrow e^+ + n$)~\cite{Vogel:1999zy}.
Energy of the antineutrino~(\enu) is estimated using the energy of the detected prompt signal~(\ep) which is generated by the positron and its annihilation gammas. An approximate relationship ${E}_{\bar\nu} \approx {E}_{p} +0.8~\text{MeV}$.
All previous measurements have focused on \ep$<$8~MeV because signals in the higher-energy region are rare and often contaminated by cosmogenic backgrounds.
Despite the rarity, high-energy reactor antineutrinos represent a background for future measurements, such as the diffuse supernova neutrino background that is expected to permeate the Universe~\cite{Beacom:2003nk}.

High-energy reactor antineutrinos are likely generated by only a handful of short-lived $\beta$-decay nuclei with high end-point energies ($Q_{\beta}$), such as $^{88,90}$Br and $^{94,96,98}$Rb.
Daughters of these high-$Q_{\beta}$ nuclei have complex deexcitation pathways that may have been incorrectly determined by measurements suffering from low efficiency for associated $\gamma$ rays (the ''pandemonium effect''~\cite{Hardy:1977suw}).
Recent measurements using the total absorption spectroscopy revealed the substantial impact of pandemonium effects in comparatively lower $Q_{\beta}$ isotopes on the calculation of reactor antineutrino spectra~\cite{Fallot:2012jv,IGISOL:2015ifm,Valencia:2016rlr,Rice:2017kfj,Guadilla:2019aiq,Estienne:2019ujo,PhysRevLett.119.052503}, indicating the likelihood of similar issues at higher $Q_{\beta}$~\cite{Valencia:2016rlr}.
In this sense, direct measurements of high-energy antineutrinos can provide a valuable new perspective for nuclear data validations relevant well beyond the bounds of neutrino physics~\cite{Algora:2020mhh,bib:IAEA,Schmidt:2020ngx}.

This Letter reports the measurement of reactor antineutrinos with \ep$>$8~MeV using a multivariate analysis at Daya Bay.
This is the first detailed analysis of reactor-produced antineutrinos in this energy region, yielding an unfolded antineutrino energy spectrum which provides a more reliable data-based prediction than existing theoretical methods.

The Daya Bay experiment studies antineutrinos from the Daya Bay nuclear power complex, which hosts six commercial pressurized-water reactor cores (each with 2.9~GW maximum thermal power).
Eight identically designed antineutrino detectors (ADs) are deployed in two near experimental halls (EH1 and EH2, each containing two ADs) and in a far experimental hall (EH3, containing four ADs).
Details about the experiment are described in Refs.~\cite{An:2015qga,An:2016ses}.
IBD interaction candidates are selected following the criteria similar to ``selection A'' described in Ref.~\cite{An:2016ses}.
While in previous analyses the veto of IBD candidates following cosmic muon signals with high reconstructed energy ($E_{\mu}^{\rm rec}>2.5$~GeV) was set to 1~s to suppress backgrounds from cosmogenic isotopes, the veto time is shortened to 1~ms to include these background-rich samples.
This change allows exploitation of background properties to reliably disentangle signal from background.
In the latest dataset taken in 1958 calendar days, we collected about 4 million IBD candidates at three experimental halls but only about 9000 candidates with \ep~between 8 and 12~MeV.

\begin{figure}
    \centering
    \includegraphics[width=\columnwidth]{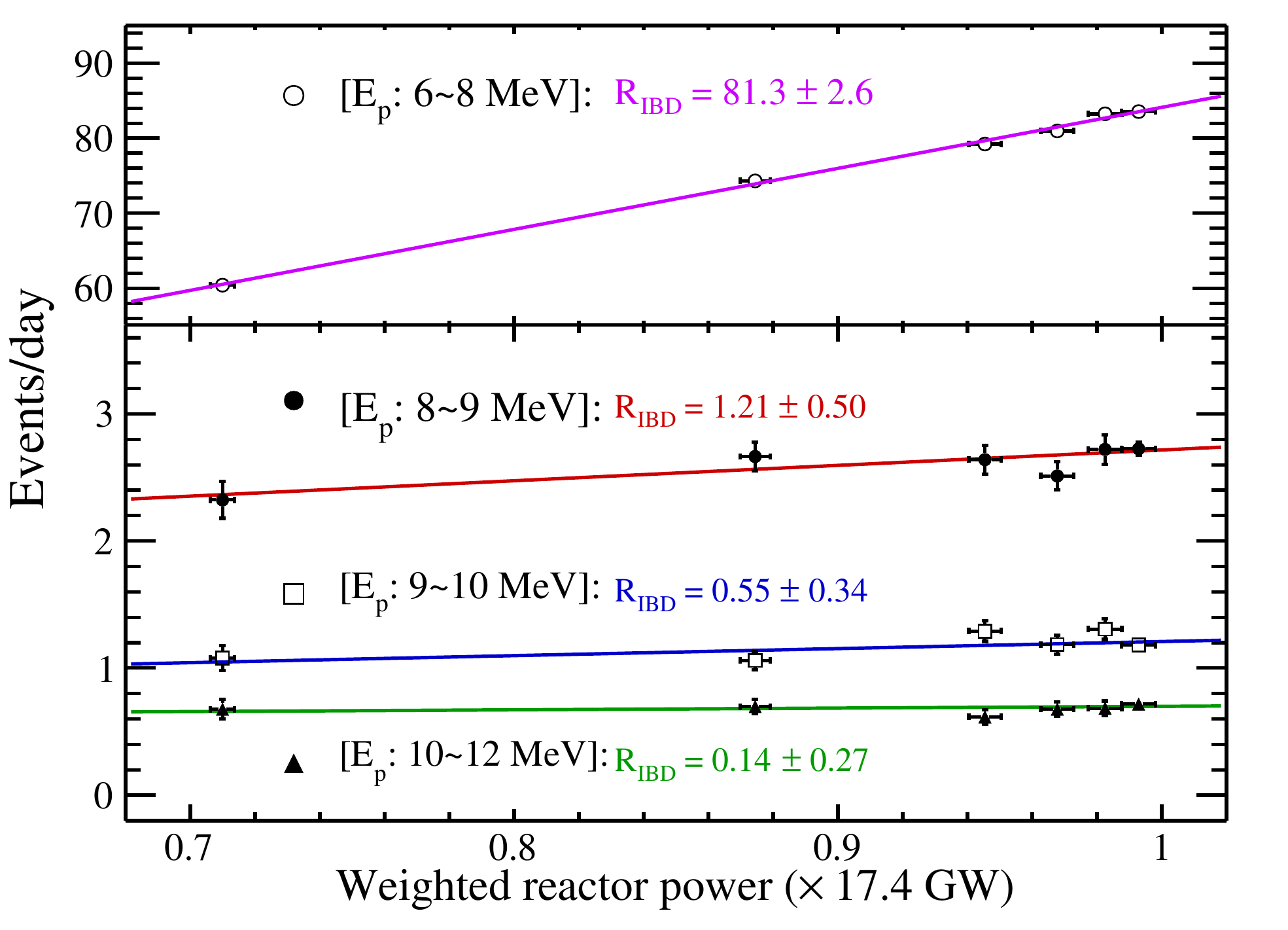}
    \caption{ Relationship between IBD rates without background subtract in four \ep~bins and the weighted reactor power. The vertical error bars are statistical uncertainties and the horizontal ones are systematic uncertainties. The relationship is fitted with a linear function. Scopes which represent rates of IBD signals are also shown. }
    \label{fig:ReactorPower_EHAll}
\end{figure}

Figure~\ref{fig:ReactorPower_EHAll} displays the IBD candidate rate $R$ as a function of the weighted reactor power, ${P}_{\rm reactor}$, for four \ep~bins. $R$ is the sum of all detectors without background subtraction. For each detector, the ${P}_{\rm reactor}$ is defined as
\begin{equation}
\label{Eq:Peffective}
{P}_{\rm reactor} = \frac{R_{\rm RealPower}}{R_{\rm FullPower}} \times 17.4~{\rm GW_{th}},
\end{equation}
where $R_{\rm RealPower}$ is the predicted IBD rate using the reactor information provided by the company and $R_{\rm FullPower}$ is the predicted rate assuming that all reactors were working with full power.
It is found that the event rate drops by more than a factor of 20 for the 8-12~MeV energy region compared to that of 6-8~MeV energy region, suggesting a much larger background to signal ratio in the high-\ep~region.
By decomposing $R$ into two parts, $R=R_{\rm IBD}{P}_{\rm reactor}+R_{\rm bkg}$, where $R_{\rm IBD}$ represents the IBD reaction rate per unit of reactor power, and $R_{\rm bkg}$ represents the background rate which is uncorrelated with ${P}_{\rm reactor}$,
a strong correlation between $R_{\rm IBD}$ and ${P}_{\rm reactor}$ is observed for \ep~in 6 to 8 MeV that is inconsistent with a background-only hypothesis at over 30$\sigma$.
By contrast, the significance of the correlation decreases to $<$2.5$\sigma$ above 8 MeV, primarily due to the much larger background to signal ratio. This figure illustrates significant challenges in extracting the IBD events with \ep$>$ 8~MeV.

The main backgrounds with \ep~in 8 to 12~MeV are from muon decays, cosmogenic fast neutrons, and cosmogenic isotope decays.
Other contributions, such as atmospheric neutrinos, are negligible.
A probability distribution function~(PDF) is constructed for each event:
\begin{equation}
\label{Eq:TotalPDF}
F(\boldsymbol{r};\boldsymbol{\Delta t},z,w)=\sum_p r_p f_p(\boldsymbol{\Delta t})h_p(z)k_p(w),
\end{equation}
where $p$ represents the event types (IBD, cosmogenic isotopes, or fast neutron), $r_p$ is the ratio of the number of type-$p$ events over the total event number in each \ep~bin, and $\boldsymbol{r}$ represents the vector of $r_p$.
$f(\boldsymbol{\Delta t})$ is the expected distribution of time difference $\boldsymbol{\Delta t}$ between the IBD candidate and its preceding muon events in the same detector (the vector of $\Delta t$ for each of eight muon categories described below),
$h(z)$ is the distribution of vertical vertex coordinates $z$ of the prompt signal,
and $k(w)$ is the event distribution of the weighted reactor power $w$ when the event occurred.
$\boldsymbol{r}$ is determined by minimizing a $\chi^2$ constructed as
\begin{equation}
\label{Eq:Chi2}
\chi^{2}(\boldsymbol{r}) =-2\sum\left[ \log F(\boldsymbol{r};\boldsymbol{\Delta t},z,w) \right]+g(\boldsymbol{\epsilon}),
\end{equation}
where $\Sigma$ denotes the sum over all events and $g(\boldsymbol{\epsilon})$ constrains the nuisance parameters describing $f_p(\boldsymbol{\Delta t})$, $h_p(z)$ and $k_p(w)$ utilizing information described below.

If a cosmic-ray muon stops in the detector, its energy deposit and the subsequent Michel electron could mimic an IBD event.
Muons were tagged with an efficiency larger than 99.7\% by the two water Cherenkov detectors surrounding ADs. However, the efficiency decreased after 2017 because of an increase of nonfunctional photomultipliers in the Cherenkov detectors, resulting in the appearance of muon decay background.
The background was negligible for previous neutrino oscillation analyses. In this analysis it was found that a $z>1.7$~m cut applied on the prompt signal can reject more than 99.8\% of this background, because stopping muons with less than 12~MeV energy deposits concentrated in the top of AD. This background is not considered in Eq.~\ref{Eq:TotalPDF} due to the negligible contribution after applying the cut.

Neutrons generated by cosmic muon interactions outside ADs could penetrate into ADs and mimic IBD reactions. They are dubbed "fast neutrons".
Because of the production mechanism, the majority of fast neutrons have downward momentum and deposit energies near top of ADs.
The $z$ distribution fast neutrons for 8$<$\ep$<$12~MeV is inferred from an almost pure fast neutron sample with 12$<$\ep$<$20~MeV.
The $z$ distribution of antineutrinos is uniform and determined using the IBD-enriched sample with $2<$\ep$<8$~MeV that has less than 0.1\% fast neutron contamination.
Rates of fast neutrons obtained using the vertex information are consistent with an alternative method based on IBD-like events coincident with muons identified using only the outer water Cherenkov detector~\cite{An:2016ses}.

The cosmogenic production of $^9$Li and $^8$He (referred as $^9$Li in later text) with subsequent $\beta-n$ decay or a coincidence of two $^{12}$B $\beta$ decays can generate a correlated pair of signals nearly identical to IBDs.
Since cosmogenic isotopes are produced by muons penetrating through the whole detector, their $z$ vertex distributions are found to be consistent with the uniform distribution of IBDs.
The $\Delta t$ distribution is described by $f(\Delta t)=\kappa \cdot e^{-\kappa \Delta t}$, where $\kappa = R_{\mu}$ for muon-uncorrelated events such as IBDs and $\kappa = R_{\mu}+\frac{n}{\tau}$ for muon-correlated events.
Here, $R_{\mu}$ is the muon rate, and $\tau$ is the life of the isotope. Details are found in Ref~\cite{Wen:2006hx}, Fig.~26 in Ref.~\cite{An:2016ses}, and Fig.~2 in Ref.~\cite{Adey:2018zwh}. The factor $n$ is 1 for $^{9}$Li and 2 for the accidental coincidence of two $^{12}$B $\beta$ decays.
The different $\Delta t$ dependencies facilitate an effective determination of amounts of cosmogenic isotopes.
In addition, production of unstable isotopes is usually associated with energetic particle showers attributed to muons~\cite{KamLAND:2009zwo}, which often include spallation neutrons.
Thus, muons are divided into eight categories: four $E^{\rm rec}_\mu$ bins, and each bin with or without accompanying neutrons.

The cosmogenic background rates can be effectively determined from the $\Delta t$ distributions except in the near halls (EH1 and EH2) for the lowest muon-energy bin ($E^{\rm rec}_\mu < 1$~GeV) without accompanying neutrons.
The high muon rates in these two cases result in indistinguishable $\Delta t $ distributions for cosmogenic isotope backgrounds and muon-uncorrelated events.
Previous analyses in Daya Bay studied the relation between isotope and neutron productions~\cite{An:2016ses,Adey:2018zwh}: $\varepsilon_{hi}=Y_{hi}^{n}/Y_{hi}^{\rm all}$, where $Y_{hi}^{\rm all}$ is the total isotope yield from muons in the $i^{th}$ $E_{\mu}^{\rm rec}$ bin in experimental hall $h$, and $Y_{hi}^{n}$ is the yield with accompanying neutrons.
Though EH3 is situated deeper underground than EH1 or EH2, $\varepsilon$ is similar in all three halls.
Thus the fitted $Y^{n}$ for $E^{\rm rec}_\mu < 1$~GeV can be combined with the determination of $\varepsilon$ in EH3 to estimate $Y^{\rm all}$ in EH1 and EH2.

An independent method was used to cross-check $^9$Li yields from low-energy muons in EH1 and EH2.
A full GEANT4-based~\cite{GEANT4:2002zbu} simulation of muons was performed.
For each muon track through Gd-doped liquid scintillator, energies from steps with large energy losses are summed to yield a "shower energy".
The shower energy distribution $\boldsymbol{S}_{hi}^{\rm sh}$ is obtained for muons in the $i^{\rm th}$ $E_{\mu}^{\rm rec}$ range.
The relationship between the known $^9$Li yields from data ($Y^{\rm all}_{hi}$) and $\boldsymbol{S}^{\rm sh}_{hi}$ is built by: $Y^{\rm all}_{hi}=\sum_j \boldsymbol{S}^{\rm sh}_{hij}\times \boldsymbol{Q}_{j}$, where $\boldsymbol{Q}$ represents the $^9$Li yield from each muon in the $j^{th}$ shower energy bin.
When extracting the $\boldsymbol{Q}$ based on $Y^{\rm all}_{hi}$ and $\boldsymbol{S}^{\rm sh}_{hi}$, an unfolding technique is utilized to suppress the amplified fluctuation~\cite{AlmazanMolina:2020jlh}.
Finally, with the determined $\boldsymbol{Q}$, the $^9$Li yield from the lowest muon-energy bin is found to account for 18\% (13\%) of the total yield in EH1 (EH2), with 50\% (55\%) relative uncertainty.
These predictions are consistent with the results calculated based on $\varepsilon$ and are used as constraints in the pull term of Eq.~\ref{Eq:Chi2}.

\begin{figure}
    \centering
    \includegraphics[width=\columnwidth]{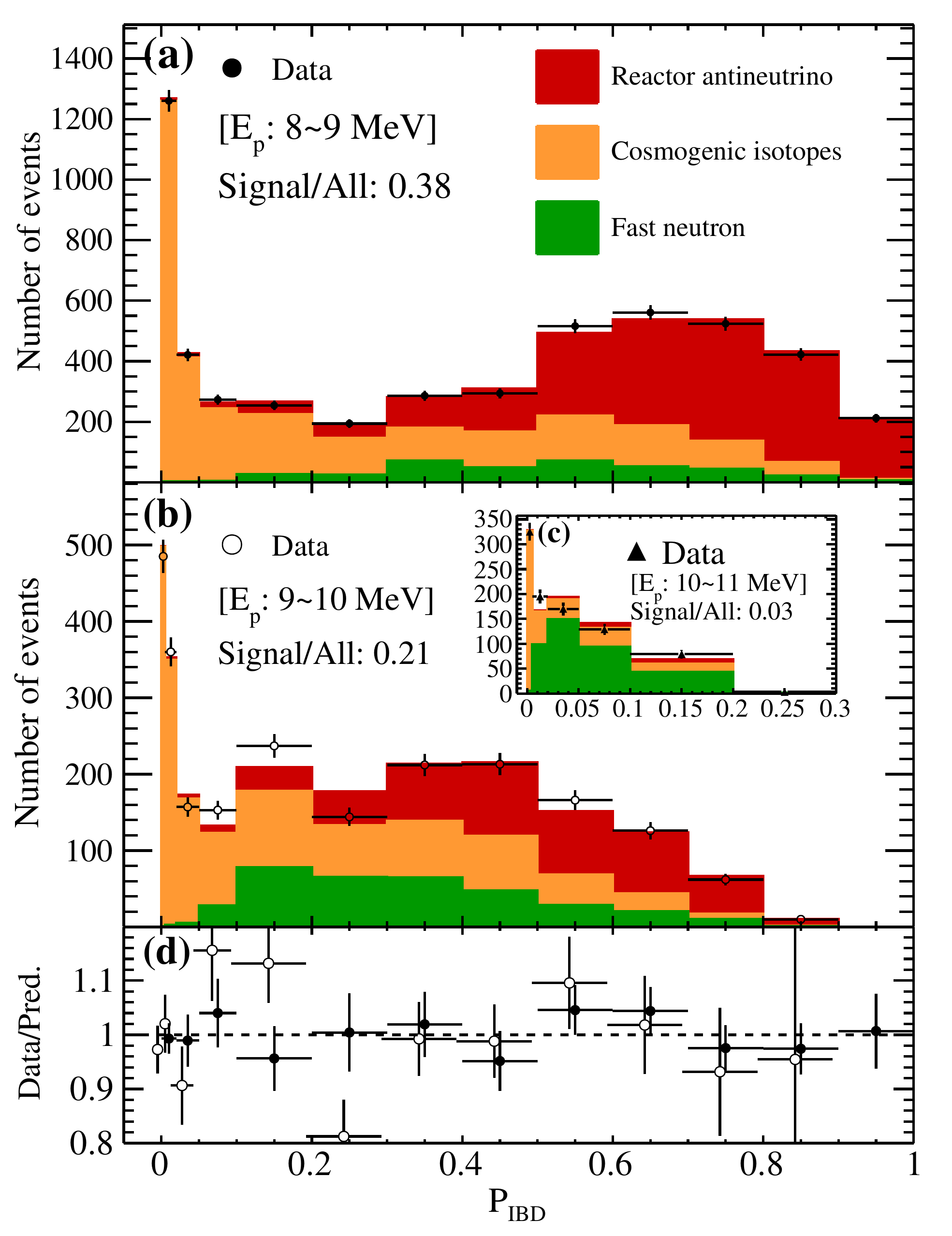}
    \caption{ (a-c) Distributions of probabilities of being an IBD signal based on the best-fit results for different event types in different \ep~bins ((a): 8-9, (b): 9-10, and (c): 10-11~MeV). The filled histograms show the expected distributions which are calculated using toy Monte Carlo datasets generated from the PDFs of the fitting model.
    (d) Ratios of the distributions between data and predictions in (a) and (b).
    The error bars are statistical only.
    The data points for \ep~in the 9-10~MeV bin are shifted slightly left for visual clarity. }
    \label{fig:ProbIBD_EHAll}
\end{figure}

After the minimization of Eq.~\ref{Eq:Chi2}, using the best-fit values of $\boldsymbol{r}$, the probability of being an IBD signal is calculated for each event:
\begin{equation}
\label{Eq:Prob_IBD}
{P}_{\rm IBD}=\frac{r_{\rm IBD} f_{\rm IBD}(\boldsymbol{\Delta t})h_{\rm IBD}(z)k_{\rm IBD}(w)}{F(\boldsymbol{r};\boldsymbol{\Delta t},z,w)}.
\end{equation}
The predictions of ${P}_{\rm IBD}$ for different event types are calculated using toy Monte Carlo datasets generated from the PDFs in Eq.~\ref{Eq:TotalPDF}.
As illustrated in Fig.~\ref{fig:ProbIBD_EHAll}, the reactor antineutrinos can be statistically separated and the differentiating capability decreases as the prompt energy increases.
The ${P}_{\rm IBD}$ distributions from data and predictions are consistent within statistical uncertainty, which demonstrates good agreement between data and model.

It is found that cosmogenic isotopes are the dominant background for \ep$<$9.5~MeV and fast neutrons for \ep$>$9.5~MeV.
As a sanity check, removing reactor power information from the fitter leads to a similar result because the reactors are almost always operating at full power.
Removing the $\Delta t$ information or reconstructed vertex information yields consistent results with significantly larger uncertainties.
Analysis with the data of only one EH also shows consistent results within 1$\sigma$.
The uncertainty of the IBD yields for \ep$>$8~MeV is dominated by systematic uncertainties from the background estimation.
Uncertainty from cosmogenic isotopes takes up to $\sim$60\% of the total uncertainty for \ep~in 8 to 9.5~MeV, while the uncertainty from fast neutrons accounts for more than 70\% for \ep$>$10~MeV.

The above analysis has extracted the antineutrino yield for \ep$>$8~MeV. An unfolding technique is needed to remove the detector response and convert the \ep~spectrum to the \enu~spectrum. This requires the \ep~spectrum in lower-energy regions. The prompt energy spectrum below 8~MeV has been updated with the new $^9$Li background estimation in this analysis. The result is consistent with those released in previous publications~\cite{Adey:2019ywk,DayaBay:2021dqj} within uncertainties.

\begin{figure}
    \centering
    \includegraphics[width=\columnwidth]{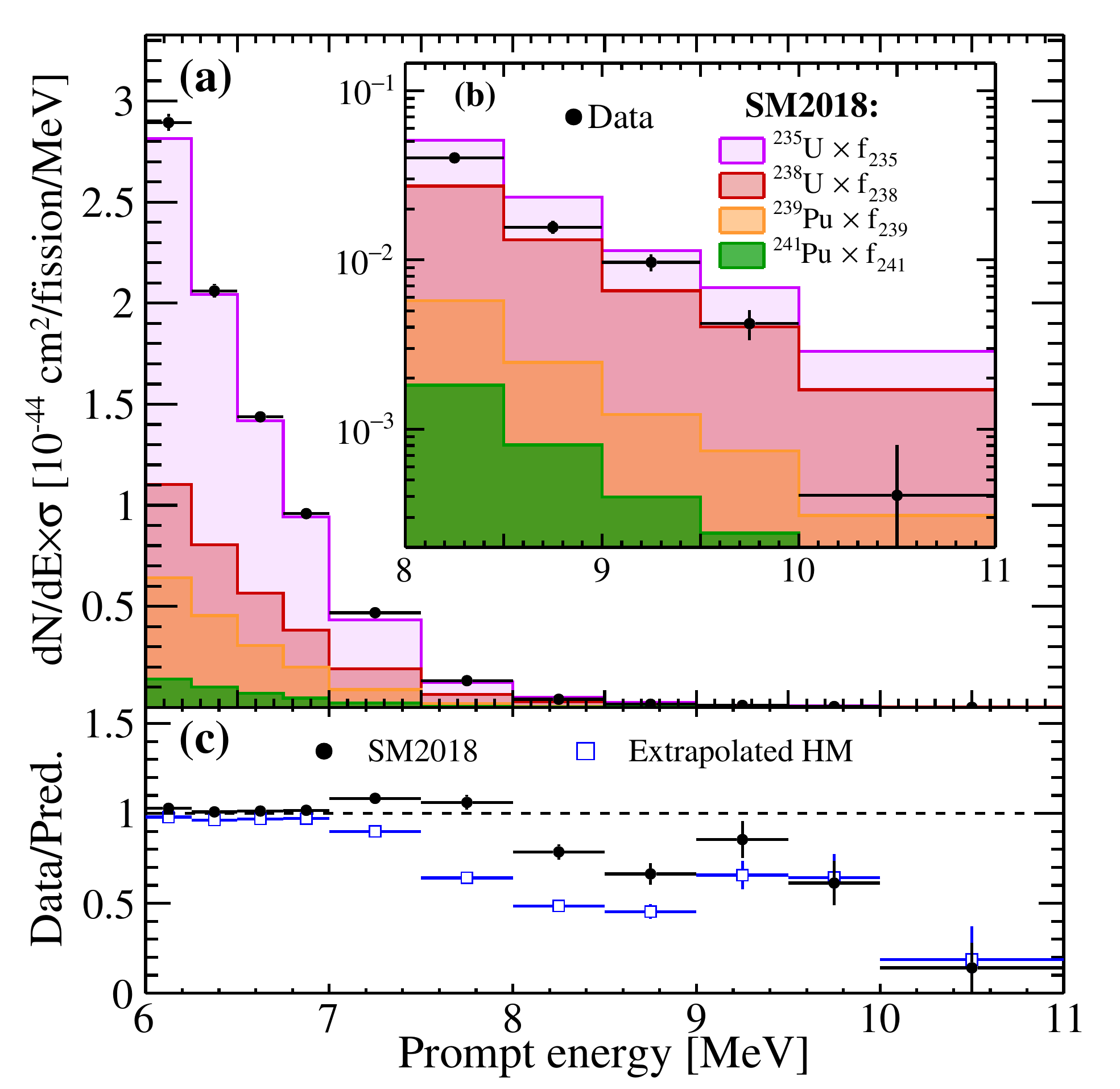}
    \caption{
    (a) Measured prompt energy spectrum compared with the prediction from the SM2018 model. The components from different isotopes ($^{235}$U, $^{238}$U, $^{239}$Pu, and $^{241}$Pu) are shown for fission fractions (0.564, 0.076, 0.304, and 0.056 respectively).
    (b) A enlarged plot of (a) above 8~MeV with logarithmic vertical scales.
    (c) Ratio of the measurement over the prediction from SM2018 and the extrapolated Huber-Mueller (HM) model.
    The HM model is not valid above 8 MeV as described in the text.
    The error bars in the data points are the square root of the diagonal elements of the covariance matrix, including both the statistical uncertainty and systematic uncertainty. The result above 11 MeV is not shown due to the larger than 100\% uncertainty.
    }
    \label{fig:MeasuredSpectrumHERAv}
\end{figure}

The prompt energy spectrum per fission above 6~MeV and the comparison with theoretical predictions are shown in Fig.~\ref{fig:MeasuredSpectrumHERAv}.
An extended neutrino spectrum for \enu$>$10~MeV was kindly provided by the authors of a recent summation model (SM2018)~\cite{Estienne:2019ujo} for the comparison.
The measured IBD yield is 3\% larger than the SM2018 prediction for \ep~in 6-8~MeV (similar to the conclusion from Ref.~\cite{DayaBay:2021dqj}) but is 29\% smaller than the prediction for \ep~in 8-11~MeV.
Interestingly, for \ep$>$10~MeV, while a nonzero signal rate of greater than 5$\sigma$ statistical significance is predicted by the SM2018 model, the significance is only 1$\sigma$ for the data.
The large deficit in the observed high-energy IBD yields is consistent with the presence of pandemonium-affected decay data for high-$Q_{\beta}$ isotopes such as $^{94,96,98}$Rb~\cite{Valencia:2016rlr,Dwyer:2014eka} in the SM2018 summation prediction, which would result in overstated feeding to low-lying states and an overprediction of high-energy antineutrinos.
In this case, overprediction of high-energy antineutrinos would be accompanied by underprediction of beta-delayed neutron release by these isotopes~\cite{PhysRev.55.664,INDC(NDS)-0599,Birch:2014raf}.

The measured high-energy IBD yields were also compared to predictions derived from the Huber-Mueller beta conversion model~\cite{Huber:2011wv,Mueller:2011nm}.
A polynomial extrapolation was used to obtain the predictions for $E_\nu>$8~MeV~\cite{Huber:2011wv}.
The extrapolated result is larger than the measurement by 30\% or more for \ep$>$7.5~MeV, and gives worse agreement with data than the SM2018 model.
For this reason, we recommend not using this extrapolation in future high-energy reactor antineutrino studies.

A data-based reactor antineutrino energy spectrum is determined with an unfolding technique, similar to the earlier analysis in the low-energy region~\cite{DayaBay:2021dqj}.
We constructed a fitter by removing the detector response based on
\begin{equation}
\chi^{2}=(\boldsymbol{P}-\boldsymbol{M}) \textbf{Cov}^{-1}(\boldsymbol{P}-\boldsymbol{M})^{T}.
\end{equation}
Here the $\boldsymbol{M}$ and $\textbf{Cov}$ are the measured prompt energy spectrum per fission above 6~MeV and its covariance matrix, respectively.
The prediction of the prompt energy spectrum per fission ($\boldsymbol{P}$) is calculated by: $\boldsymbol{P}=\boldsymbol{R} \boldsymbol{S}^{\mathrm{fit}}$,
where $\boldsymbol{R}$ is the detector response matrix at Daya Bay, which maps the predicted antineutrino energy spectrum ($\boldsymbol{S}^{\mathrm{fit}}$) to prompt energy.
Here $\boldsymbol{S}_{i}^{\mathrm{fit}}=\boldsymbol{S}_{i}^{\rm init} \times \boldsymbol{\eta}_{i}$, with free parameter $\boldsymbol{\eta}_i$ on the initial values ($\boldsymbol{S}_{i}^{\rm init}$) in $i^{\rm th}$ energy bin.
The starting point of the antineutrino energy spectrum in the unfolding is set at 6~MeV to ensure that resolution-induced feedup of 6-7~MeV antineutrinos into the high-energy region is properly accounted for.

\begin{figure}
    \centering
    \includegraphics[width=\columnwidth]{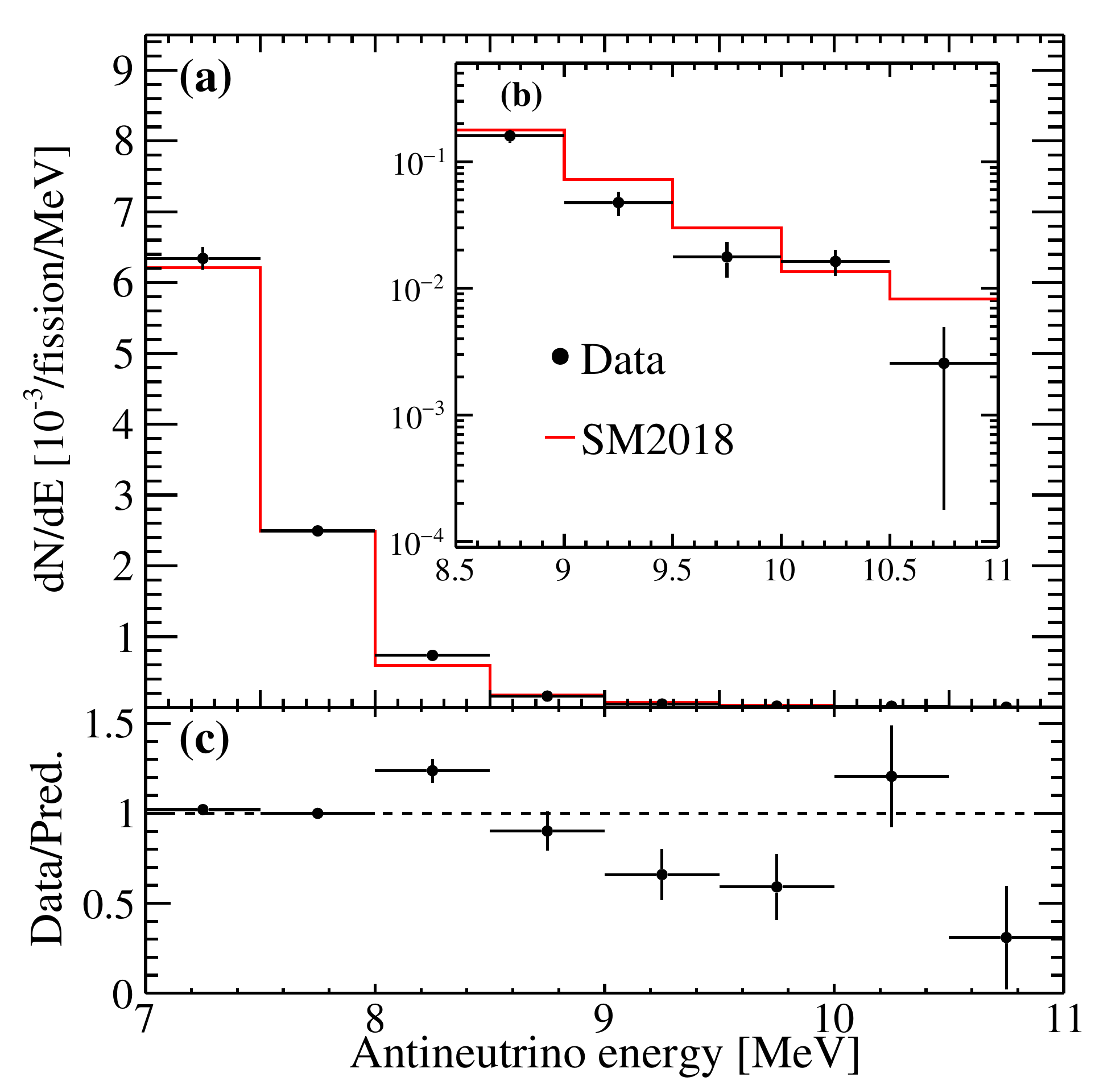}
    \caption{
    (a) Reactor antineutrino energy spectrum above 7~MeV and its comparison with the prediction from SM2018. The IBD cross section~\cite{Vogel:1999zy} has been removed.
    The fission fractions of $^{235}$U, $^{238}$U, $^{239}$Pu, and $^{241}$Pu are 0.564, 0.076, 0.304, and 0.056 respectively.
    The central values have larger deviation from SM2018 compared with Fig.~\ref{fig:MeasuredSpectrumHERAv} due to the amplified statistical fluctuations with bin-to-bin anticorrelation in the unfolding process~\cite{DayaBay:2021dqj}.
    (b) A enlarged plot of (a) above 8.5~MeV with a logarithmic vertical scale.
    (c) Ratio of the unfolded antineutrino energy spectrum over the prediction from SM2018.
    }
    \label{fig:UnfoldedSpectrumHERAv}
\end{figure}

The unfolded antineutrino energy spectrum is shown in Fig.~\ref{fig:UnfoldedSpectrumHERAv}.
In the unfolding process, the postfit prediction ($\boldsymbol{P}$) is the same as the measurement ($\boldsymbol{M}$), with the best-fit $\chi^2$ value equal to 0.
Mathematically, this method is equivalent to the matrix inversion method
, but this procedure has the advantage of enabling a variety of statistical tests.
While this method could suffer from amplified statistical fluctuations and big bin-to-bin anticorrelation in the unfolding process~\cite{DayaBay:2021dqj}, these problems are mitigated by the large bin widths used in this analysis.
Assuming no reactor antineutrinos above specific energies (10, 10.5, or 11~MeV) in $\boldsymbol{S}^{\mathrm{fit}}$, the best-fit $\chi^2$ values (38.3, 1.6, and 0.03, respectively) are obtained.
Therefore, the significance of our result in rejecting the hypothesis of no reactor antineutrinos above 10~MeV is 6.2$\sigma$.
Above 10.5~MeV, the ability to reject the background-only hypothesis is marginal.
The prompt energy spectrum, unfolded antineutrino energy spectrum and their covariance matrices are included in  Supplemental Material.

In summary, the Daya Bay experiment has determined the prompt IBD energy spectrum up to 11~MeV.
A 29\% difference in IBD rate in the prompt energy region of 8-11~MeV is found compared with a recent summation model.
An antineutrino energy spectrum is then obtained from these data using an unfolding procedure that removes the detector response effects.
The hypothesis of no reactor antineutrinos with energy above 10~MeV is rejected with a significance of 6.2$\sigma$.
For the first time, this work extends the energy region of reactor antineutrinos above 10~MeV by direct measurement.
The combination of very high statistics and low cosmogenic backgrounds of Daya Bay suggests that the precision of this measurement is unlikely to be surpassed in the foreseeable future.

We thank Magali Estienne and Muriel Fallot for their generosity to extend their summation model in Ref.~\cite{Estienne:2019ujo} above 10~MeV and their helpful discussions.

The Daya Bay experiment is supported in part by the Ministry of Science and Technology of China, the U.S. Department of Energy, the Chinese Academy of Sciences, the CAS Center for Excellence in Particle Physics, the National Natural Science Foundation of China, the Guangdong provincial government, the Shenzhen municipal government, the China General Nuclear Power Group, the Research Grants Council of the Hong Kong Special Administrative Region of China, the Ministry of Education in Taiwan, the U.S. National Science Foundation, the Ministry of Education, Youth, and Sports of the Czech Republic, the Charles University Research Centre UNCE, the Joint Institute of Nuclear Research in Dubna, Russia, the National Commission of Scientific and Technological Research of Chile, We acknowledge Yellow River Engineering Consulting Co., Ltd., and China Railway 15th Bureau Group Co., Ltd., for building the underground laboratory. We are grateful for the ongoing cooperation from the China Guangdong Nuclear Power Group and China Light \& Power Company.

\bibliography{references}

\end{document}

%% file: PhysRev.tex
\newcommand{\IHEP}{\affiliation{Institute~of~High~Energy~Physics, Beijing}}
\newcommand{\Wisconsin}{\affiliation{University~of~Wisconsin, Madison, Wisconsin 53706}}
\newcommand{\Yale}{\affiliation{Wright~Laboratory and Department~of~Physics, Yale~University, New~Haven, Connecticut 06520}} 
\newcommand{\BNL}{\affiliation{Brookhaven~National~Laboratory, Upton, New York 11973}}
\newcommand{\NTU}{\affiliation{Department of Physics, National~Taiwan~University, Taipei}}
\newcommand{\NUU}{\affiliation{National~United~University, Miao-Li}}
\newcommand{\Dubna}{\affiliation{Joint~Institute~for~Nuclear~Research, Dubna, Moscow~Region}}
\newcommand{\CalTech}{\affiliation{California~Institute~of~Technology, Pasadena, California 91125}}
\newcommand{\CUHK}{\affiliation{Chinese~University~of~Hong~Kong, Hong~Kong}}
\newcommand{\NCTU}{\affiliation{Institute~of~Physics, National~Chiao-Tung~University, Hsinchu}}
\newcommand{\NJU}{\affiliation{Nanjing~University, Nanjing}}
\newcommand{\TsingHua}{\affiliation{Department~of~Engineering~Physics, Tsinghua~University, Beijing}}
\newcommand{\SZU}{\affiliation{Shenzhen~University, Shenzhen}}
\newcommand{\NCEPU}{\affiliation{North~China~Electric~Power~University, Beijing}}
\newcommand{\Siena}{\affiliation{Siena~College, Loudonville, New York  12211}}
\newcommand{\IIT}{\affiliation{Department of Physics, Illinois~Institute~of~Technology, Chicago, Illinois  60616}}
\newcommand{\LBNL}{\affiliation{Lawrence~Berkeley~National~Laboratory, Berkeley, California 94720}}
\newcommand{\UIUC}{\affiliation{Department of Physics, University~of~Illinois~at~Urbana-Champaign, Urbana, Illinois 61801}}
\newcommand{\SJTU}{\affiliation{Department of Physics and Astronomy, Shanghai Jiao Tong University, Shanghai Laboratory for Particle Physics and Cosmology, Shanghai}}
\newcommand{\BNU}{\affiliation{Beijing~Normal~University, Beijing}}
\newcommand{\WM}{\affiliation{College~of~William~and~Mary, Williamsburg, Virginia  23187}}
\newcommand{\Princeton}{\affiliation{Joseph Henry Laboratories, Princeton~University, Princeton, New~Jersey 08544}}
\newcommand{\VirginiaTech}{\affiliation{Center for Neutrino Physics, Virginia~Tech, Blacksburg, Virginia  24061}}
\newcommand{\CIAE}{\affiliation{China~Institute~of~Atomic~Energy, Beijing}}
\newcommand{\SDU}{\affiliation{Shandong~University, Jinan}}
\newcommand{\NanKai}{\affiliation{School of Physics, Nankai~University, Tianjin}}
\newcommand{\UC}{\affiliation{Department of Physics, University~of~Cincinnati, Cincinnati, Ohio 45221}}
\newcommand{\DGUT}{\affiliation{Dongguan~University~of~Technology, Dongguan}}
\newcommand{\XJTU}{\affiliation{Department of Nuclear Science and Technology, School of Energy and Power Engineering, Xi'an Jiaotong University, Xi'an}}
\newcommand{\UCB}{\affiliation{Department of Physics, University~of~California, Berkeley, California  94720}}
\newcommand{\HKU}{\affiliation{Department of Physics, The~University~of~Hong~Kong, Pokfulam, Hong~Kong}}
\newcommand{\Charles}{\affiliation{Charles~University, Faculty~of~Mathematics~and~Physics, Prague}} 
\newcommand{\USTC}{\affiliation{University~of~Science~and~Technology~of~China, Hefei}}
\newcommand{\TempleUniversity}{\affiliation{Department~of~Physics, College~of~Science~and~Technology, Temple~University, Philadelphia, Pennsylvania  19122}}
\newcommand{\CGNPG}{\affiliation{China General Nuclear Power Group, Shenzhen}}
\newcommand{\NUDT}{\affiliation{College of Electronic Science and Engineering, National University of Defense Technology, Changsha}} 
\newcommand{\IowaState}{\affiliation{Iowa~State~University, Ames, Iowa  50011}}
\newcommand{\ZSU}{\affiliation{Sun Yat-Sen (Zhongshan) University, Guangzhou}}
\newcommand{\CQU}{\affiliation{Chongqing University, Chongqing}} 
\newcommand{\BCC}{\altaffiliation[Now at ]{Department of Chemistry and Chemical Technology, Bronx Community College, Bronx, New York  10453}} 

\newcommand{\UCI}{\affiliation{Department of Physics and Astronomy, University of California, Irvine, California 92697}} 
\newcommand{\GXU}{\affiliation{Guangxi University, No.100 Daxue East Road, Nanning}} 
\author{F.~P.~An}\ZSU
\author{W.~D.~Bai}\ZSU
\author{A.~B.~Balantekin}\Wisconsin
\author{M.~Bishai}\BNL
\author{S.~Blyth}\NTU
\author{G.~F.~Cao}\IHEP
\author{J.~Cao}\IHEP
\author{J.~F.~Chang}\IHEP
\author{Y.~Chang}\NUU
\author{H.~S.~Chen}\IHEP
% ...
